\documentclass{cernrep} 
\usepackage{texnames}
\usepackage[T1]{fontenc}
\usepackage{graphicx}
\usepackage{amssymb}
\usepackage{epstopdf}
\usepackage{epstopdf}

\begin{document}
\title{Error bars for distributions of numbers of events}
 
\author{R. Aggarwal$^{1,2}$, A.~Caldwell$^2$}

\institute{$^1$ Panjab University, Chandigarh, India (ritu.aggarwal1@gmail.com)\\
$^2$ Max Planck Institute for Physics, Munich, Germany (caldwell@mpp.mpg.de)}

\maketitle 

\begin{abstract}
The common practice for displaying error bars on distributions of numbers of events is confusing and can lead to incorrect conclusions.  A proposal is made for a different style of presentation that more directly indicates the level of agreement between expectations and observations.
\end{abstract}

\section{Introduction}
\begin{figure}[htbp] 
   \centering
   \includegraphics[width=6in]{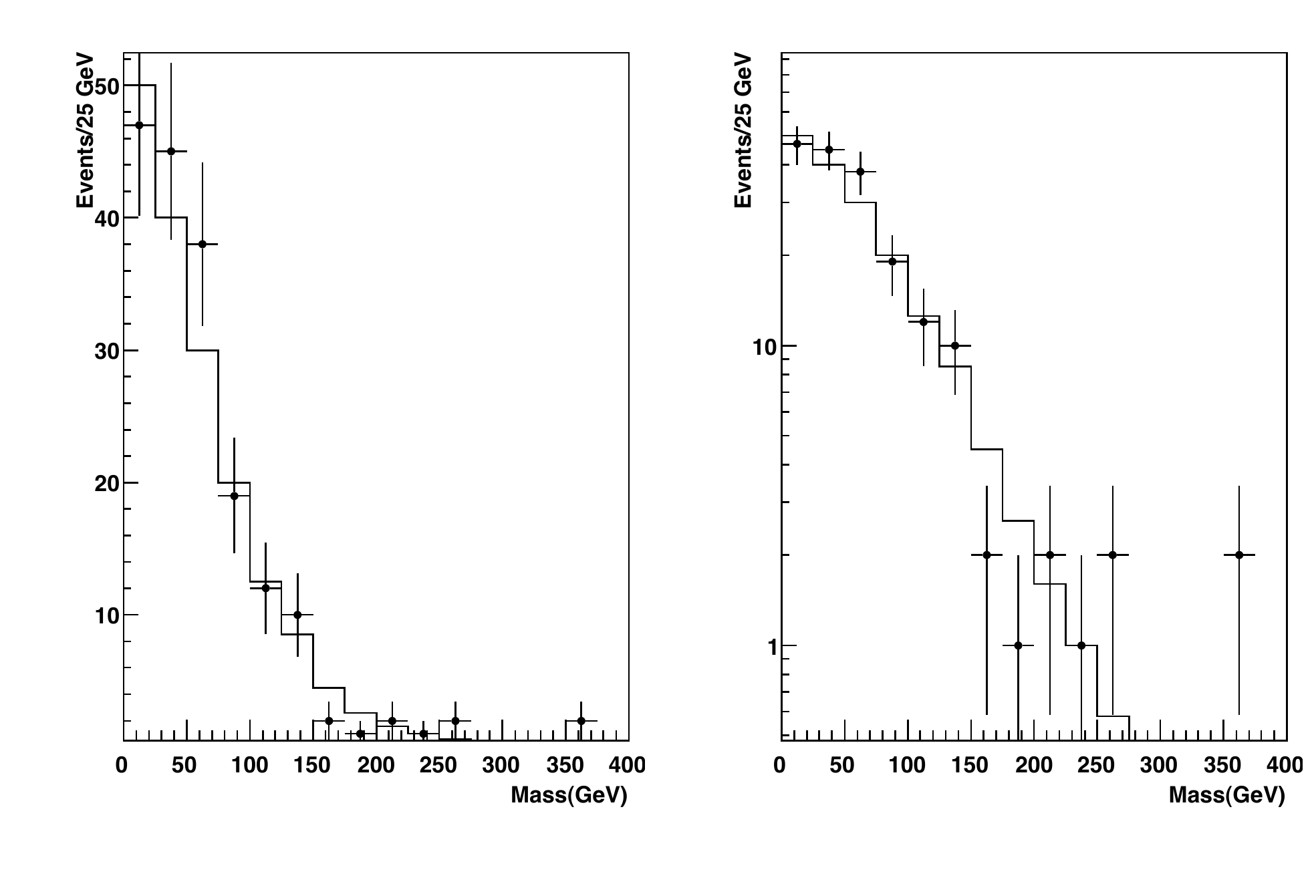}    
      \caption{A typical presentation of data - here event counts as a function of mass in $25$~GeV intervals.  The same data are plotted twice: left - linear scale; right - logarithmic scale. The error bars cover the range $(o-\sqrt{o},o+\sqrt{o})$, where $o$ is the number of observed events.  The histogram gives the expectations (means of Poisson distributions). }
   \label{fig:bad}
\end{figure}

Symmetric error bars centered on the observed number of events can be highly misleading and in practice often generates confusion.  A typical example\footnote{The data is artificial and invented for the purposes of this note.} of such a data presentation is given in Fig.~\ref{fig:bad}.

 There are three problems with this standard style of presentation:
\begin{itemize}
\item

First of all, there is no uncertainty on the number of observed events. We certainly do not mean that there is a high probability that we had $2.3$ rather than $2$ events in the 7$^{\rm th}$ bin in the plot.  Actually, the error bar is intended to represent the uncertainty on a different quantity-the uncertainty on the mean of an assumed underlying Poisson distribution. The probability distribution for this mean given an observed number of events $o$, $P(\theta|o)$, can be quite asymmetric, and different choices can be made regarding what to plot as summary (e.g., the mode, the mean or the median) of $P(\theta|o)$.
\item
The second problem arises with the length of the error bar.  This is routinely taken as $\pm \sqrt{o}$ , motivated by the Poisson result that the variance is equal to the mean, so that the error bar should cover $\pm 1~\sigma$, or $68$~\% probability for possible values of $\theta$.  However, the probability range covered by this definition of the error bar varies dramatically as $o \rightarrow 0$ and the probability above and below the point is highly asymmetric. 
Usually no error bar is plotted when $0$ events are measured, although this measurement also yields information on possible values of $\theta$. These problems are occasionally avoided by using asymmetric error bars, usually covering the central 68~\% of the probability  from the cumulative of $P(\theta|o)$, but this is still the exception rather than the rule in experimental particle physics.

\item
A third problem occurs when data are compared to expectations, as in Fig.~\ref{fig:bad}, and the error bar is used to determine if the observed number of events represents a significant deviation from the expectation.  The error bar on the plot often gives the completely wrong information in this case, since the relevant probability is the probability that the expectation could have yielded the number of observed events, not the probability that the observed number of events could have fluctuated to the expectation. For example, in the next-to-last bin in Fig.~\ref{fig:bad}, the expectation is $0.011$ and two events are observed. It is VERY wrong to conclude that we have slightly more than a $1~\sigma$ discrepancy in this bin.
\end{itemize}

A proposal for an alternative presentation is given here.  We focus on the case where we are comparing observations to predictions and the fluctuations can be modeled with a Poisson distribution.  We start with the simplest case - that predictions are available with negligible uncertainty.  We then show how to include the uncertainty due to predictions based on finite Monte Carlo event sets and due to systematic uncertainties.

\section{Negligible uncertainty on the prediction}
We start with the simplest case - the expectations (means of Poisson distributions) are known with very small uncertainty, and we want to compare the observed data to these expectations.  The plot should give us an indication of whether the observations are within reasonable statistical fluctuations of the expectation; i.e., the plot should give the user an indication of how rare a particular observed number of events $o$ was expected to be given a Poisson distribution with mean number of events $\nu$.  The probability distribution for $o$ is 
\begin{equation}
P(o|\nu)=\frac{e^{-\nu}\nu^{o}}{o!} \;\; .
\end{equation}
The most probable value for $o$ (mode of the probability distribution) is given by 
\begin{equation}
o^*=\lfloor \nu \rfloor \;\; ;
\end{equation}
i.e., the largest integer not greater than $\nu$.

Two choices can be made for the probability intervals to display:
\begin{enumerate}
\item The central interval, defined for a probability density $P(x)$ with possible range for the parameter $\{x_{\rm min}, x_{\rm max}\}$
$$\alpha/2=\int_{x_{\rm min}}^{x_1} P(x) dx=\int_{x_{2}}^{x_{\rm max}} P(x) dx \;\;.$$
The central interval is $\{x_1,x_2\}$ and contains probability $1-\alpha$.  In our case, we have a discrete probability distribution $P(o|\nu)$ and the equality generally cannot be satisfied.  We take instead
\begin{eqnarray}
\label{eq:central}
o_1 &=& \sup_{o \in \mathbb{Z}}\{ \sum_{i=0}^o P(i|\nu) \le \alpha/2\} + 1 \\
o_2 &=& \inf_{o \in \mathbb{Z}} \{ \sum_{i=o}^{\infty} P(i|\nu) \le \alpha/2\}  - 1 \;\; .
\end{eqnarray}
If $P(o=0|\nu)>\alpha/2$, then we take $o_1=0$.

We define the set of observations which fall into the central $1-\alpha$ probability interval\footnote{Note that $1-\alpha$ is the minimum probability covered and that the set generally covers a larger probability.}  as
$$\mathcal{O}_{1-\alpha}^{\rm C} = \{o_1,o_1+1, ..., o_2\}$$
and we display these values of $o$.  Different colors can be used to represent different $1-\alpha$ probability ranges.
For example, if we take $\nu=3.\bar{3}$, then we find the values given in Table~\ref{tab:Poisson}.  If we choose $1-\alpha=0.68$, then our definitions give
$$\mathcal{O}_{0.68}^{\rm C} = \{2,3,4,5\} $$
whereas
$$\mathcal{O}_{0.95}^{\rm C} = \{0,1,2,3,4,5,6,7\} $$
and
$$\mathcal{O}_{0.999}^{\rm C} = \{0,1,2,3,4,5,6,7,8,9,10,11\} \;\; .$$

\begin{table}[htdp]
\caption{Values of $o$, the probability to observe such a value given $\nu=3.\bar{3}$ and the cumulative probability, rounded to four decimal places. The fourth column gives the rank in terms of probability - i.e., the order in which this value of $o$ is used in calculating the smallest set $\mathcal{O}_{1-\alpha}^S$, and the last column gives the cumulative probability summed according to the rank.}
\begin{center}
\begin{tabular}{c|c|c|c|c}
$o$ & $P(o|\nu)$ & $F(o|\nu)$ & $R$ & $F_R(o|\nu)$ \\
\hline
\hline
0 & 0.0357 & 0.0357 & 7 & 0.9468\\
1 & 0.1189 & 0.1546 & 5 & 0.8431 \\
2 & 0.1982 & 0.3528 & 2 & 0.4184 \\
3 & 0.2202 & 0.5730 & 1 & 0.2202 \\
4 & 0.1835 & 0.7565 & 3 & 0.6019\\
5 & 0.1223 & 0.8788 & 4 & 0.7242\\
6 & 0.0680 & 0.9468 & 6 & 0.9111 \\
7 & 0.0324 & 0.9792 & 8 & 0.9792 \\
8 & 0.0135 & 0.9927 & 9 & 0.9927 \\
9 & 0.0050 & 0.9976 &10 & 0.9976 \\
10 & 0.0017 & 0.9993 & 11 & 0.9993\\
11 & 0.0005 & 0.9998 & 12 & 0.9998 \\
12 & 0.0001 & 1.0000 & 13 & 1.0000 \\
\hline
\end{tabular}
\end{center}
\label{tab:Poisson}
\end{table}%

\item The second option for the probability interval is to use the smallest interval containing a given probability.  In the case of a unimodal continuous distribution, we can write the condition as  
$$1-\alpha = \int_{x_{1}}^{x_{2}} P(x) dx \;\; {\rm and} \;\; P(x_1)=P(x_2)\;\; .$$
For our discrete case, the set making up the smallest interval containing probability at least $1-\alpha$,   $\mathcal{O}_{1-\alpha}^{S}$, is defined by the following algorithm

\begin{enumerate}
\item
Start with $\mathcal{O}^S_{1-\alpha}=\{o^*\}$.   If 
$P(o^*|\nu)\geq 1- \alpha$, then we are done.  An example where this requirement is fulfilled for $1-\alpha=0.68$ is  ($\nu=0.001, o^*=0, \mathcal{O}_{0.68}^S=\{ 0 \}$).
\item If $P(o^*|\nu)< 1-\alpha$, then we need to add the next most probable number of observations, which in the unimodal case is either $o^*+1$ or $o^*-1$.  Assume that $P(o^*+1|\nu) > P(o^*-1|\nu)$.  Then, we would extend our set to  $\mathcal{O}_{1-\alpha}^S=\{o^*, o^*+1\}$ and check again whether $P(\mathcal{O}_{1-\alpha}^S)\geq 1-\alpha$.  We would continue to add members to the set $\mathcal{O}_{1-\alpha}^S$ until this condition is met, always taking the highest probability of the remaining possible observations.

In the special cases that two observations have exactly the same probability (when $\nu$ takes on an integer value, $P(o=\nu)=P(o=\nu-1)$), then both values should be taken in the set.

\end{enumerate}
If we consider the example given in Table~\ref{tab:Poisson}, then using the cumulative according to rank, $F_R$, we find
$$\mathcal{O}_{0.68}^{\rm S} = \{2,3,4,5\} $$
whereas
$$\mathcal{O}_{0.95}^{\rm S} = \{0,1,2,3,4,5,6,7\} $$
and
$$\mathcal{O}_{0.999}^{\rm S} = \{0,1,2,3,4,5,6,7,8,9,10\} \;\; .$$\
The same sets are found as for the central interval for $1-\alpha=0.68$ and $0.95$, but is smaller for $0.999$.
\end{enumerate}

If the predicted distribution has a large mean ($\nu> 50$, say), then we can use the Gaussian approximation.  In this case, we take the minimal symmetric range around $o^*$ such that 
$$\int_{o^*-(n+0.5)}^{o^*+(n+0.5)} \frac{1}{\sqrt{2 \pi \nu}} e^{-\frac{(x-o^*)^2}{2\nu}} dx\geq1- \alpha$$
and define $O_{\alpha}=\{o^*-n,...,o^*+n\}$.  This gives both the central interval as well as the minimal interval.

\begin{figure}[htbp] 
   \centering
   \includegraphics[width=3in]{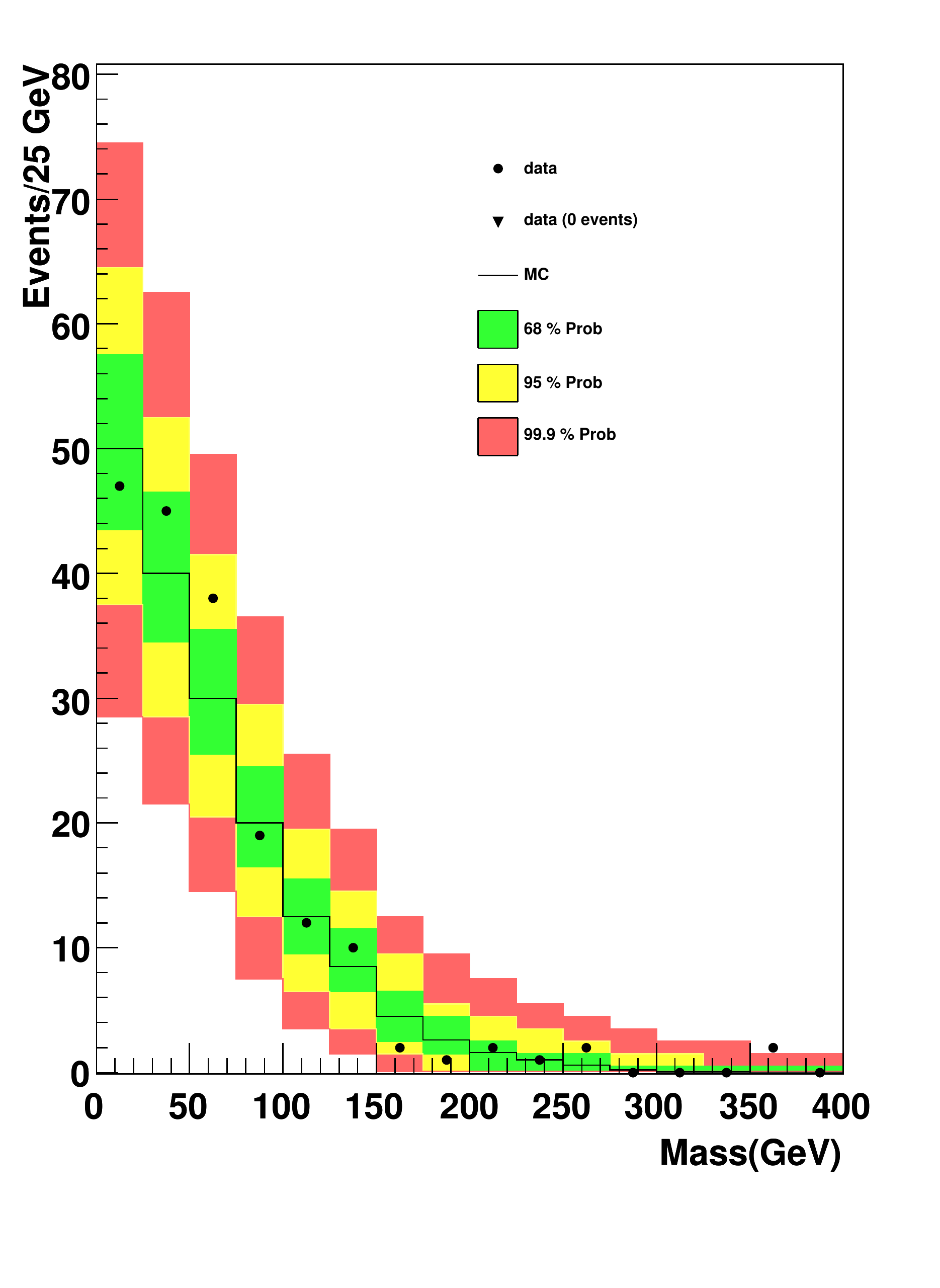}       \includegraphics[width=3in]{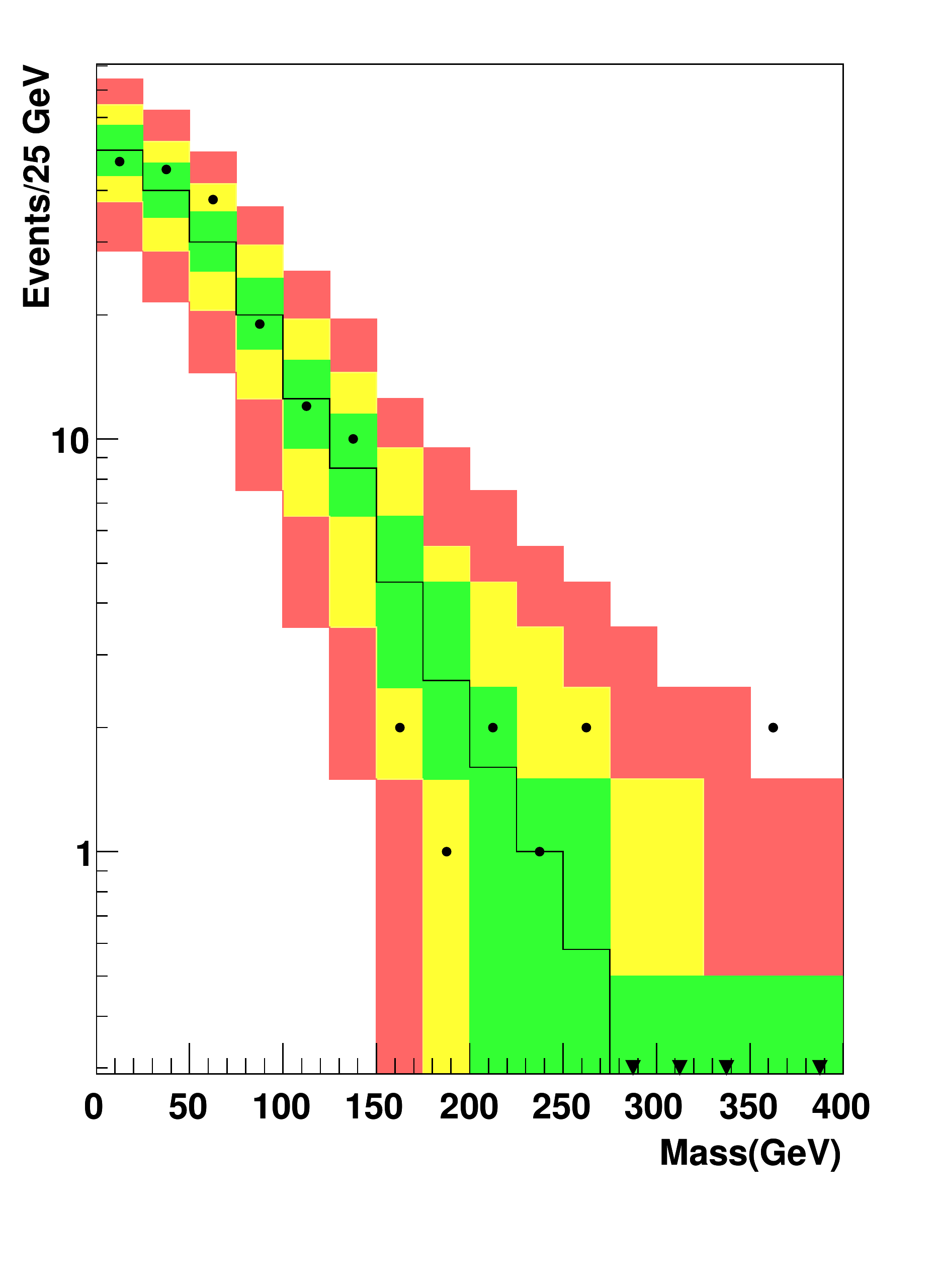}    
      \caption{The proposed style of presentation for the same data as shown in Fig.~\ref{fig:bad}. The shaded bands represent different probability intervals for the observed number of events: green is for $1-\alpha=0.68$; yellow is for  $1-\alpha=0.95$ and red is for  $1-\alpha=0.999$.  The bands use the smallest interval containing at least this probability, and extend to the next 1/2 integer value (see text).  For the logarithmic scale plot on the right, the data are shown as a solid triangle if $o=0$. }
   \label{fig:good}
\end{figure}

As an example for the procedures defined here,  Fig.~\ref{fig:good} shows the same distribution of observed number of events as a function of invariant mass as was shown in Fig.~\ref{fig:bad}.  Three different probability intervals are shown, corresponding to $1-\alpha=0.68$, $1-\alpha=0.95$, and $1-\alpha=0.999$ and follow the definition of the smallest interval.  Note that the bands are extended beyond the integer values in the set by $0.5$ for clarity of presentation.  E.g., the band containing the set $\{2,3,4,5\}$ is drawn from $1.5 \rightarrow 5.5$.  The smallest $1-\alpha$ color is chosen if more than one $1-\alpha$ set contain the same set members (e.g., this occurs when the set $\{0\}$ contains $>95$~\% probability as in the last bins in the plot). The color scheme is meant to be suggestive.  Observed event counts outside the shaded bands should indicate unlikely results.

\section{Predictions based on Monte Carlo with finite statistics}
We now extend the prescription for cases where the prediction is uncertain.  The first case we consider is that the prediction is based on a Monte Carlo which has non-negligible statistical uncertainties. We derive the probability distribution for the expected number of events $o$ given that a MC set (consisting possibly of different components) gives $n$ events and the MC normalization factor is $s$ (defined such that the MC prediction is divided by the factor $s$ when calculating the expected mean for the data).  The factor $s$ is initially taken to be known exactly.

The process of assigning Monte Carlo events to bins indicates the use of the multinomial distribution for calculations (if we generate a fixed number of events rather than a fixed luminosity). We assume here that the probability for an event to populate any given bin is small so that Poisson statistics can be used as a valid approximation.  Assume our Monte Carlo sample has resulted in $n$ events (in a bin of interest).  We now need to determine the expected mean for the data sample and the probability distribution for this mean.  We use $\lambda$ for the mean of the MC distribution in the bin, and $\nu$ for the mean expected for the observations.  Applying Bayes' Theorem~\cite{ref:Bayes} and taking the Jeffreys' prior\cite{ref:Jeffreys} on the mean for the MC, $\lambda$, 
$$ P_0(\lambda) \propto \frac{1}{\sqrt{\lambda}}$$
the pdf for $\lambda$ is
$$P(\lambda|n) = \frac{4^n n!}{\sqrt{\pi} (2n)!} \lambda^{n-1/2} e^{-\lambda}$$
which leads to
\begin{eqnarray*}
\rm{E}[\lambda]&=&n+1/2 \;\;\;\; {\rm and}\\
\lambda^* &=& n-1/2 \;\; .
\end{eqnarray*}

\noindent For scaling the prediction to the mean expected for the data $(\nu=\lambda/s)$, we have:
\begin{eqnarray*}
P(\nu|n,s)&=& P(\lambda|n) d\lambda/d\nu \\
               &=& \frac{4^n n!}{\sqrt{\pi} (2n)!}  s(s\nu)^{n-1/2} e^{-s\nu}
\end{eqnarray*}
For the distribution of data events, we use the Law of Total Probability~\cite{ref:Law}:
\begin{eqnarray*}
P(o|n,s) &=& \int P(o|\nu)P(\nu|n,s)d\nu \\
P(o|n,s) &=& \frac{s^{n+1/2}4^n n!}{\sqrt{\pi}o!(2n)!}\int \nu^{o+n-1/2}  e^{-(s+1)\nu} d\nu 
\end{eqnarray*}
The integral gives
$$\frac{[2(o+n)]!\sqrt{\pi}}{4^{n+o}(1+s)^{o+n+1/2}(n+o)!}$$
so
\begin{equation}
\label{eq:MC1}
P(o|n,s) = \frac{s^{n+1/2}}{(1+s)^{o+n+1/2}}\frac{n![2(o+n)]!}{4^o o! (2n)! (n+o)!} \;\; .
\end{equation}
and
\begin{eqnarray*}
{\rm E}[o] &=& \frac{n+1/2}{s} \;\; . \\
\end{eqnarray*}

\noindent For $o=0$, we have
\begin{equation}
\label{eq:MC2}
P(o=0|n,s) = \left( \frac{s}{1+s}\right)^{n+1/2} \;\; .
\end{equation}
The probability for the succeeding values of $o$ can then be easily calculated as
$$
P(o+1|n,s)=P(o|n,s)\cdot \frac{(2n+2o+2)(2n+2o+1)}{4(o+1)(1+s)(n+o+1)} \;\; .
$$

We would now use these probabilities of $o$ for the procedure described in the previous section rather than the Poisson distribution for $P(o|\nu)$.

\begin{table}[htdp]
\caption{Values of $o$, the probability to observe such a value given $n=10$ and $s=3$ and the cumulative probability, rounded to four decimal places. The fourth column gives the rank in terms of probability - i.e., the order in which this value of $o$ is used in calculating the smallest set $\mathcal{O}_{1-\alpha}^S$, and the last column gives the cumulative probability summed according to the rank.}
\begin{center}
\begin{tabular}{c|c|c|c|c}
$o$ & $P(o|n,s)$ & $F(o|n,s)$ & $R$ & $F_R(o|n,s)$ \\
\hline
\hline
0 & 0.0488 & 0.0488 & 7 & 0.9072\\
1 & 0.1280 & 0.1768 & 4 & 0.6654\\
2 & 0.1840 & 0.3608 & 2 & 0.3757 \\
3 & 0.1917 & 0.5525 & 1 & 0.1917 \\
4 & 0.1617 & 0.7142 & 3 & 0.5374\\
5 & 0.1173 & 0.8315 & 5 & 0.7827\\
6 & 0.0757 & 0.9072 & 6 & 0.8584 \\
7 & 0.0446 & 0.9519 & 8 & 0.9519 \\
8 & 0.0244 & 0.9763 & 9 & 0.9763 \\
9 & 0.0125 & 0.9888 &10 & 0.9888 \\
10 & 0.0061 & 0.9949 & 11 & 0.9949\\
11 & 0.0028 & 0.9978 & 12 & 0.9978 \\
12 & 0.0013 & 0.9991 & 13 & 0.9991 \\
13 & 0.0006 & 0.9996 & 14 & 0.9996 \\
14 & 0.0002 & 0.9998 & 15 & 0.9998 \\
15 & 0.0001 & 0.9999 & 16 & 0.9999 \\
\hline
\end{tabular}
\end{center}
\label{tab:Poisson-MC}
\end{table}%

\begin{figure}[htbp] 
   \centering
      \includegraphics[width=3in]{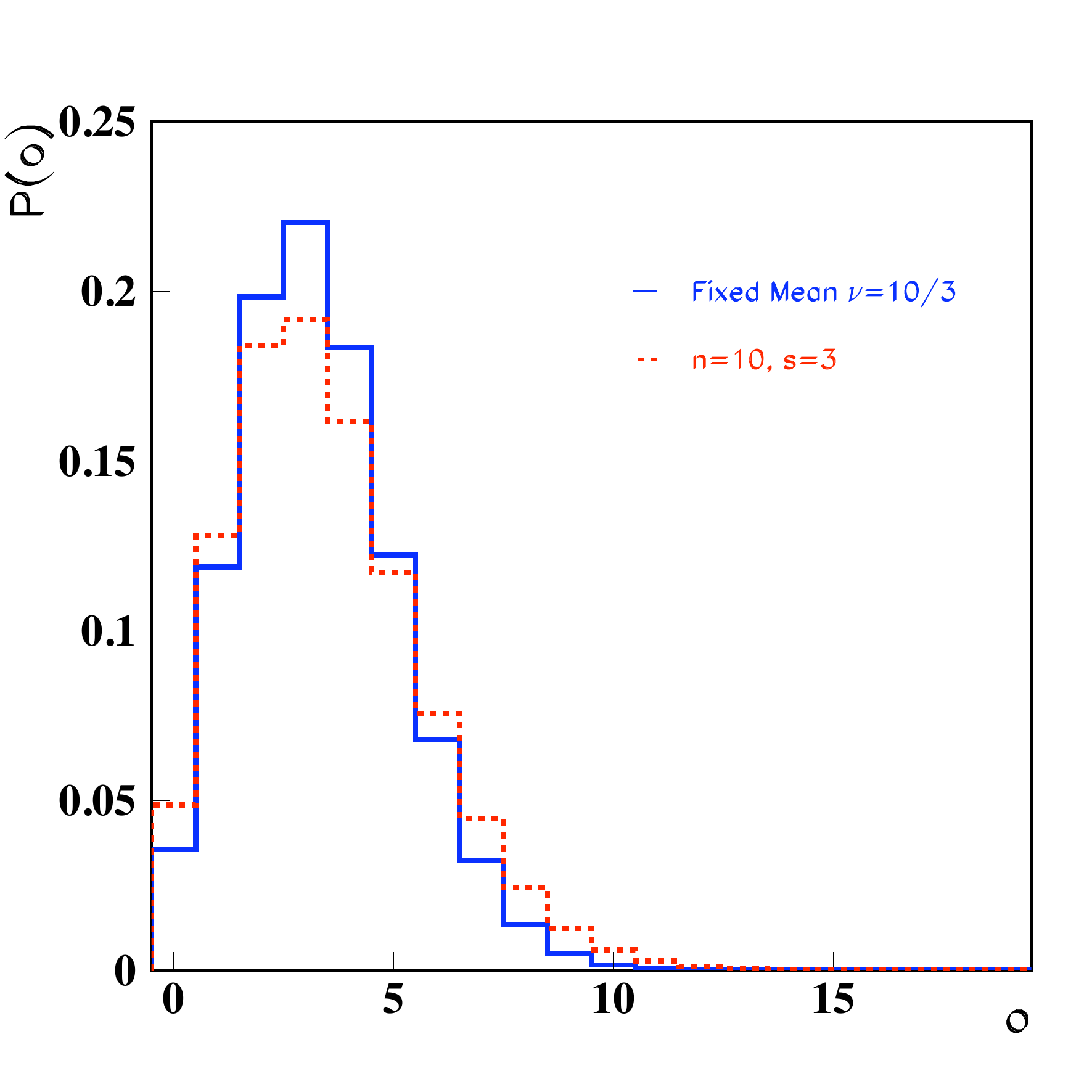}    
      \caption{Comparison of the distributions $P(o|\nu=10/3)$ and $P(o|n=10,s=3.)$.}
   \label{fig:finite}
\end{figure}

An example of the effect of finite Monte Carlo statistics on $P(o)$ is given in Fig.~\ref{fig:finite}.  As is clear, if the Monte Carlo used to derive the prediction for the number of events has an effective luminosity only a factor $3$ larger than the data, significant differences can result in probabilities for observed events. If we consider the example given in Table~\ref{tab:Poisson-MC}, where $n=10$ and $s=3$, the mean value of $\nu$ is $\bar{\nu}=3.5$ and the finite MC statistics gives slightly different results for our sets:
\begin{eqnarray*}
\mathcal{O}_{0.68}^{\rm C} = \{1,2,3,4,5,6\} & \;\;\;\;\; & 
\mathcal{O}_{0.68}^{\rm S} = \{1,2,3,4,5\} \\
\mathcal{O}_{0.95}^{\rm C} = \{0,1,2,...,8\} & \;\;\;\;\; & 
\mathcal{O}_{0.95}^{\rm S} = \{0,1,2,...,7\} \\
\mathcal{O}_{0.999}^{\rm C} = \{0,1,2,...,13\} & \;\;\;\;\; & 
\mathcal{O}_{0.999}^{\rm S} = \{0,1,2,...,12\}  \;\; .
\end{eqnarray*}
As expected, the set of possible values of $o$ has increased for a given $1-\alpha$ probability due to the extra uncertainty introduced by the finite number of Monte Carlo counts.

\section{Prediction with systematic uncertainties}
If the predictions have systematic uncertainties, then this can also be taken into account in defining the probabilities for the observed number of events.  The probability density for $\nu$ will now depend on extra quantities, and we write
$$P(\nu|n,s,\vec{\theta})$$
where $\vec{\theta}$ is a set of nuisance parameters used to describe the systematic uncertainties (e.g.,  energy scale parameters).  It may be possible to determine $P(\nu|n,s,\vec{\theta})$ directly from Monte Carlo simulations where also the systematically uncertain quantities are varied according to their belief distributions.  In this case, we would use this information and have:
$$P(o|n,s,\vec{\theta})=\int P(o|\nu) P(\nu|n,s,\vec{\theta}) d\nu \;\; .$$
In general, this integral will need to be solved numerically and the $P(o|n,s,\vec{\theta})$ then input into the prescriptions above.

In many cases, we have a fixed number of MC events and the same events are used repeatedly with different assumptions for the uncertain quantities, so that the systematic uncertainty is on the scaling parameter $s$.  In this case, we can write
$$P(\nu|n,\vec{\theta})=\int P(\nu|n,s)P(s|\vec{\theta}) ds $$
where the systematic uncertainty appears as a pdf for the scale factor. The probability distribution for $o$ is now
$$P(o|n,\vec{\theta})=\int P(o|\nu) \left[ \int P(\nu|n,s)P(s|\vec{\theta}) ds \right] d\nu \;\; .$$  

Often, we assume the belief in values for the scale factor $s$ can be modeled as a Gaussian:
$$P(s|s_0,\sigma_s)=\frac{1}{\sqrt{2\pi}\sigma_s}e^{-\frac{(s-s_0)^2}{2\sigma_s^2}}  \;\; .$$

\begin{figure}[htbp] 
   \centering
      \includegraphics[width=3in]{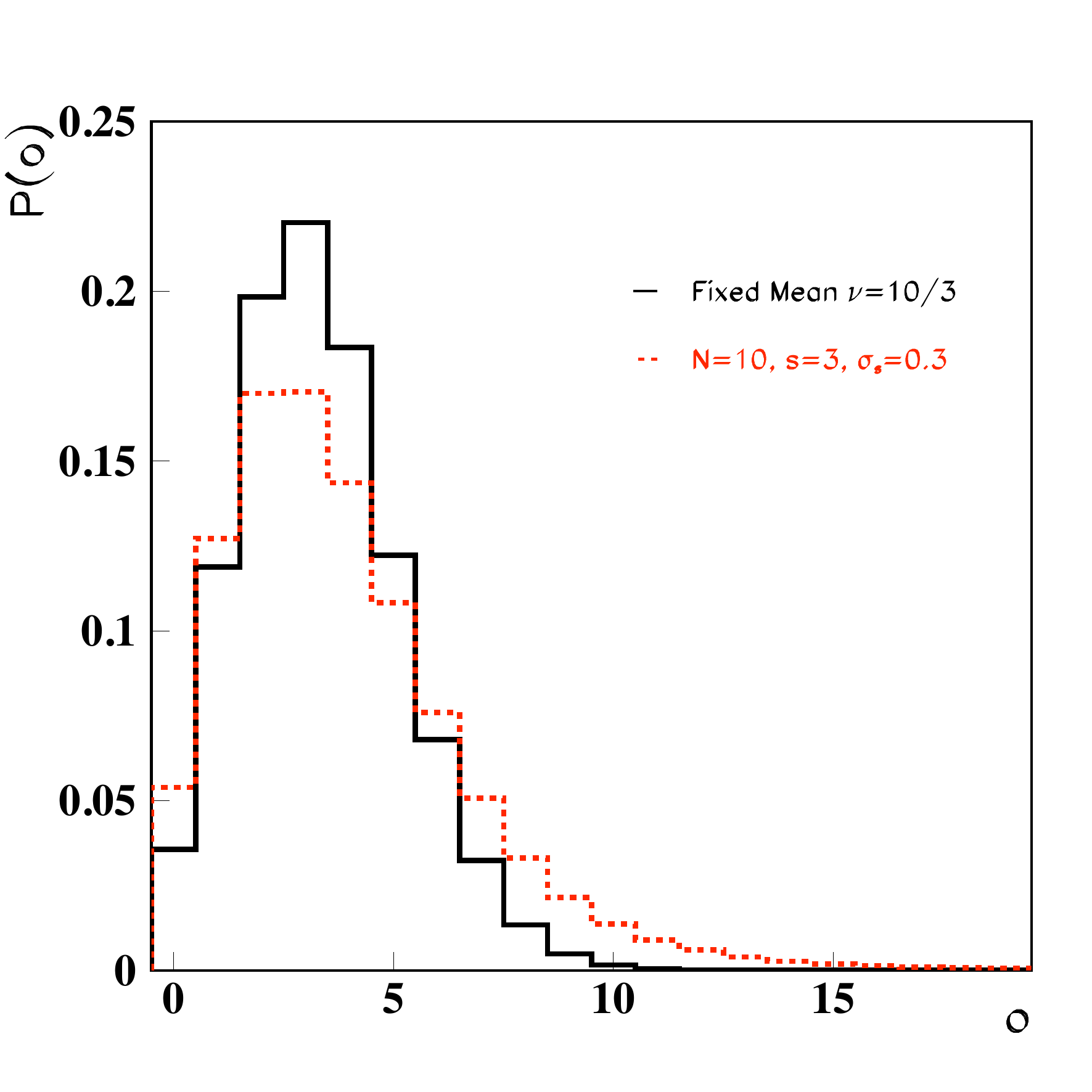}    
      \includegraphics[width=3in]{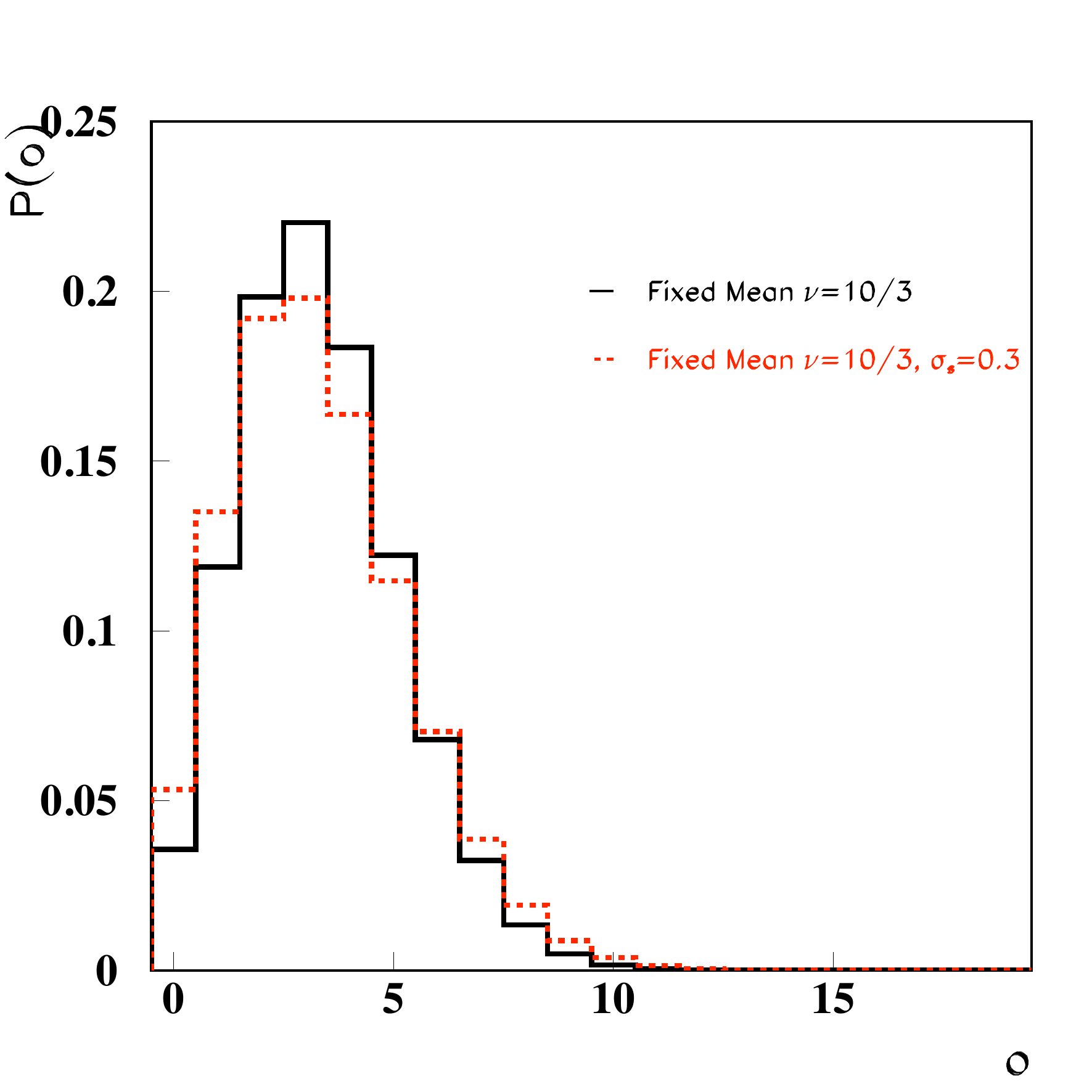}    
      \caption{Comparison of the distribution $P(o|n=10,s_0=3,\sigma_s=0.3 s_0)$ with $P(o|\nu=10/3)$ (top) and $P(o|\nu=10/3, \sigma_s=0.3s_0)$ to $P(o|\nu=10/3)$ (bottom).}
   \label{fig:syst}
\end{figure}

In this case, we would have 
\begin{equation}
\label{eq:syst}
P(o|n,s_0,\sigma_s)=\int \frac{e^{-\nu}\nu^{o}}{o!}  \left[ \int \frac{4^n n!}{\sqrt{\pi} (2n)!}  s(s\nu)^{n-1/2} e^{-s\nu}\frac{1}{\sqrt{2\pi}\sigma_s}e^{-\frac{(s-s_0)^2}{2\sigma_s^2}}ds \right] d\nu \;\; .
\end{equation}
This looks rather forbidding but can be solved numerically.  Taking our standard example, we now add to the MC statistical uncertainty a $30$~\% systematic uncertainty in the scale factor $s$ and find\footnote{note that with such a large systematic variation, the Gaussian distribution in Eq.~\ref{eq:syst} is truncated and is renormalized, leading to a shift in $E[o]$.} the results given in Table~\ref{tab:Poisson-syst}.  A graphical presentation of the effect on $P(o)$ of including systematic uncertainties on $s$ is shown in Fig.~\ref{fig:syst}.

\begin{table}[htdp]
\caption{Values of $o$, the probability to observe such a value given $n=10$ and $s=3$ and $30$~\% systematic uncertainty on $s$, together with the cumulative probability, rounded to four decimal places. The fourth column gives the rank in terms of probability - i.e., the order in which this value of $o$ is used in calculating the smallest set $\mathcal{O}_{1-\alpha}^S$, and the last column gives the cumulative probability summed according to the rank.}
\begin{center}
\begin{tabular}{c|c|c|c|c}
$o$ & $P(o|n,s)$ & $F(o|n,s)$ & $R$ & $F_R(o|n,s)$ \\
\hline
\hline
0 & 0.0539 & 0.0539 & 7 & 0.8492\\
1 & 0.1272 & 0.1811 & 4 & 0. 6111\\
2 & 0.1699 & 0.3511 & 2 & 0.3403 \\
3 & 0.1703 & 0.5214 & 1 & 0.1703 \\
4 & 0.1436 & 0.6650 & 3 & 0.4839\\
5 & 0.1083 & 0.7733 & 5 & 0.7194\\
6 & 0.0759 & 0.8492 & 6 & 0.7953 \\
7 & 0.0509 & 0.9001 & 8 & 0.9001 \\
8 & 0.0332 & 0.9333 & 9 & 0.9333 \\
9 & 0.0214 & 0.9547 &10 & 0.9547 \\
10 & 0.0138 & 0.9685 & 11 & 0.9685\\
11 & 0.0090 & 0.9776 & 12 & 0.9776 \\
12 & 0.0060 & 0.9835 & 13 & 0.9835 \\
13 & 0.0040 & 0.9876 & 14 & 0.9876 \\
14 & 0.0028 & 0.9904 & 15 & 0.9904 \\
\hline
\end{tabular}
\end{center}
\label{tab:Poisson-syst}
\end{table}%

The elements of our different sets are now given by

\begin{eqnarray*}
\mathcal{O}_{0.68}^{\rm C} = \{1,2,3,4,5,6\} & \;\;\;\;\; & 
\mathcal{O}_{0.68}^{\rm S} = \{1,2,3,4,5\} \\
\mathcal{O}_{0.95}^{\rm C} = \{0,1,2,...,11\} & \;\;\;\;\; & 
\mathcal{O}_{0.95}^{\rm S} = \{0,1,2,...,9\} \\
\mathcal{O}_{0.999}^{\rm C} = \{0,1,2,...,44\} & \;\;\;\;\; & 
\mathcal{O}_{0.999}^{\rm S} = \{0,1,2,...,33\}  \;\; .
\end{eqnarray*}

The $68$~\% interval is now different between the two definitions, and very large values of $o$ are allowed at probability $1-\alpha=0.999$, in particular for the central interval.

\section{Summary}

We have described an alternative presentation of data for cases where the aim is to judge whether an observed number of events is consistent with the model predictions.  We believe this style of presentation is more appropriate than the common one, where error bars are placed on the observed number of events.   Example code snippets for the examples described here can be requested from the authors.

\section{Acknowledgments}
The authors would like to thank Frederik Beaujean and Kevin Kr\"oninger for useful discussions.


\begin{thebibliography}{9}

\bibitem{ref:Bayes}
see, e.g., G. D'Agostini, {\it Bayesian Reasoning in Data Analysis - A Critical Introduction}, World Scientific, 2003.

\bibitem{ref:Jeffreys}
H. Jeffreys, {\it An Invariant Form for the Prior Probability in Estimation Problems}, Proceedings of the Royal Society of London. Series A, Mathematical and Physical Sciences 186 (1946) 453.

\bibitem{ref:Law}
see, e.g., D. Zwillinger and S. Kokoska, S. CRC Standard Probability and Statistics Tables and Formulae, CRC Press. Press (2000).

\end{thebibliography}
\end{document}